\documentclass[a4paper,twoside,11pt]{article}
\usepackage{ifpdf}
\usepackage{RR}
\RRNo{6216}
\usepackage{hyperref}
\usepackage{figlatex}

\usepackage{texgraphicx}
\graphicspath{{fig/}}
\usepackage{ifthen}
\usepackage[T1]{fontenc}
\usepackage{amsfonts,amsmath,amssymb,amstext,latexsym,stmaryrd}
\newcommand{\origmath}[1]{{\everymath{}\ensuremath{\everymath{}#1}}}%
\usepackage{xspace}
\usepackage{subfigure}

\usepackage{amsthm_alvin}

\newtheorem{lemma}{Lemma}
\newtheorem{theorem}{Theorem}

\newtheorem{definition}{Definition}
\newtheorem{remark}{Remark}

\let\leq=\leqslant
\let\geq=\geqslant
\let\preceq=\preccurlyeq
\let\succeq=\succcurlyeq
\def\pprec{\ensuremath{\ll}}
\def\psucc{\ensuremath{\gg}}

\let\epsilon=\varepsilon
\let\rho=\varrho
\def\R{\ensuremath{\mathbb{R}}\xspace}
\def\Rp{\ensuremath{\mathbb{R}_+}\xspace}
\newcommand\Rpn[1][n]{\ensuremath{\Rp^{#1}}\xspace}
\newcommand\Rn[1][n]{\ensuremath{\R^{#1}}\xspace}
\renewcommand\H[1][n]{\ensuremath{\mathcal{H}(\Rpn[n])}\xspace}
\def\U{\ensuremath{\mathcal{U}}\xspace}
\def\C{\ensuremath{\mathcal{C}(\Rpn)}\xspace}
\def\Pb{\ensuremath{\overline{\mathcal{P}}}\xspace}
\def\P{\ensuremath{\mathcal{P}}\xspace}

\def\ie{i.e.\xspace}
\def\eg{e.g.,\xspace}

\newcommand{\cons}[1][]{\ensuremath{\alpha\ifthenelse{\equal{#1}{}}{}{^{(#1)}}}\xspace}

\newcommand{\lowerarea}{\ensuremath{\vcenter{\hbox{\includegraphics[width=1em]{fig/lowerarea.fig}}}}\xspace}

\newsavebox{\rsbox}
\newcommand{\I}[1][\infty]{\ensuremath{\widetilde{I}_{#1}}\xspace}
\newcommand{\ISDF}{\ensuremath{I_{SDF}}\xspace}

\DeclareFixedFont{\auacc}{OT1}{phv}{m}{n}{12}   
\DeclareFixedFont{\afacc}{OT1}{phv}{m}{n}{10}   

\begin{document}

\RRdate{June 2007}

\begin{lrbox}{\rsbox}%
  \scalebox{.82}{$\log(U)\subseteq (\log(\beta(U))+\epsilon) + \lowerarea$}%
\end{lrbox}

\RRauthor{
  Arnaud Legrand \and Corinne Touati
}

\authorhead{A. Legrand \& C. Touati}
\RRtitle{Comment mesurer l'efficacité?}
\RRetitle{How to measure efficiency?}
\titlehead{How to measure efficiency?}

\RRresume{Dans le cadre de la théorie des jeux appliquée aux réseaux
  de communications, de nombreux concepts ont été proposés afin de
  mesurer l'efficacité et l'optimalité des mécanismes d'allocation de
  ressources, les plus célebres exemples étant probablement le prix de
  l'anarchie et l'index de Jain. Cependant, rares sont ceux qui ont
  cherché à étudier ces mesures et à les comparer entre elles, dans un
  contexte général. C'est ce que propose de faire cet article.}

\RRabstract{
  In the context of applied game theory in networking environments, a
  number of concepts have been proposed to measure both efficiency and
  optimality of resource allocations, the most famous certainly being
  the price of anarchy and the Jain index. Yet, very few have tried to
  question these measures and compare them one to another, in a general
  framework, which is the aim of the present article.
}

\RRkeyword{Game theory, price of anarchy, Nash equilibrium, Pareto
  optimality, Braess-like paradox.  }

\RRmotcle{Théorie des jeux, prix de l'anarchie, équilibres de Nash,
  optimalité de Pareto, paradoxes de Braess.}

\RRprojets{MESCAL}
\URRhoneAlpes
\RRtheme{\THNum}
\makeRR

\section{Introduction}

The networking community has witnessed an impressive a\-mount of work
based on applications of game theory concepts. This paper focuses here
on the ones dealing with characterizations of performance of general
policies.

We do not deal here with the choice of users utility functions. We
consider some general utilities $u$, may they represent throughput,
experienced delays\dots or any utility function and study in this paper
different allocation policies. We distinguish in particular two kinds
of policies:
\begin{itemize}
\item those who are index-based, that is to say that result on the
  optimization of a given function, as for example the Nash Bargaining
  Solution (also called proportional fairness), that maximizes the
  product of the users' utilities, or the social utility (maximizing
  their sum).
\item general policy optimization. Those do not optimize a specific
  function. The most common example being the Nash equilibrium.
\end{itemize}

While many definitions of efficiency measure can be found in the
literature, at the present day, it seems that no fully satisfactory
concept is available. The goal of this article is to present and study
various commonly used characterizations of the performance of policies.

After introducing some general notations (Section~\ref{sec:intro}), we
present some qualitative characterization of the allocations
(Section~\ref{sec:qual}): in particular the notion of Pareto
efficiency (a general notion of efficiency), of index-optimization
(that would reflect some particular property of efficient points, as
for example fairness), and Braess-like paradoxes (a particularly
non-desirable property of allocation policies). We then analyze
properties of allocations, in particular regarding continuity (which
ensures some stability of the allocation for slight changes of the
resources) and monotonicity (which ensures that an adding of resources
will always be beneficiary to the users).

Then, in Section~\ref{sec:quant}, we consider quantitative measures of
efficiency. In particular, we discuss the concepts of Jain index,
Price of anarchy (and more generally of index-optimizing based
metrics) and the recently introduced SDF (Selfishness Degradation
Factor).

\section{Notations}
\label{sec:intro}

We consider a $n$-player game, each of them having a utility function
whose values belong to \Rp. A utility set $U$ is thus a subset of
$\Rpn$. Let \H denote the set non-empty compact sets of $\Rpn$ and \C
denote the set of non-empty compact and convex sets of $\Rpn$. In the
rest of this article, we assume that \U the set of all utility sets is
either equal to \H or \C. Any negative result regarding \C also
applies to \H.

We define in this section the two kinds of allocation studied
(index-based or not) and two concepts that will turn useful for the
analytical study, namely the Hausdorff metric and some canonical
partial orders. 

\begin{definition}[Policy function]
  A policy function $\alpha:\U\to \Rpn$ is a function such that
  for all $U\in\U$, $\alpha(U)\in U$. 

  Policy functions defined on \H are said to be \emph{general
    policy function} and policy functions defined on \C are said
  to be \emph{convex policy function}.
\end{definition}

Note that in this framework, we do not consider policy optimization
that depend on previous states of the system. Such systems can occur
for instance when considering dynamic systems where Nash equilibria
adjusts to the system evolution. In the event of multiple equilibria,
the initial conditions have an impact on the convergence point.

\begin{definition}[Index-optimizing]
  An index function $f$ is a function from $\Rpn$ to $\Rp$.
  Let $f$ be an index function from $\Rpn$ to $\Rp$. A policy
  function $\alpha$ is said to be \emph{$f$-optimizing} if for all
  $U \in\U, f(\alpha(U)) = \sup_{u\in U} f(u)$.

  Index may also be called aggregation operators~\cite{Detyniecki_phd}
\end{definition}

To study the continuity of policy functions, we need a topology on
$\U$. That is why in the following, we use the classical metric on
compact sets.

\begin{definition}[Hausdorff metric]
  Considering a metric function $d$ on \Rpn, one can define the
  \emph{distance from} $x$ \emph{to} the compact $B$ as:
  \begin{equation*}
    d(x,B)=\min \{d(x,y) | y\in B\}
  \end{equation*}
  The \emph{distance from} the compact $A$ \emph{to} the compact $B$ as:
  \begin{equation*}
    d(A,B)=\max \{d(x,B) | x\in A\}
  \end{equation*}
  The \emph{Hausdorff distance} between two compacts $A$ and $B$ can
  thus be defined as:
  \begin{equation*}
    h(A,B)=\max(d(A,B),d(B,A))
  \end{equation*}
\end{definition}
$(\H,h)$ and $(\C,h)$ are complete metric spaces~\cite{topo_book} and
we can thus study the continuity of policy functions under pretty
clean conditions.

\begin{definition}[Canonical partial orders]\label{canonical}
  We consider the following orders as being canonical.
  \begin{itemize}
  \item The canonical partial $\preceq$ order on $\Rpn$ is defined by:
    \begin{equation*}
      u\preceq v \Leftrightarrow \forall k: u_k \leq v_k
    \end{equation*}
  \item The canonical partial order on $\H$ is the classical inclusion
    order: $\subseteq$.
  \end{itemize}
  The two classical strict partial order $\prec$ and $\subset$ are
  defined accordingly. 
  \begin{itemize}
  \item We also define an additional strict partial order $\pprec$ on
    $\Rpn$, namely the strict Pareto-superiority, by:
    \begin{equation*}
      u\pprec v \Leftrightarrow \forall k: u_k < v_k
    \end{equation*}
  \end{itemize}
\end{definition}

\section{Qualitative Characterizations}
\label{sec:qual}

In this section, we focus on qualitative characterizations of
performance of allocations. Of particular interest are:
\begin{itemize}
\item The notion of Pareto optimality: a concepts that define the set
  of points of $U$ that are globally optimal,
\item Index or aggregation operators: they reflect the optimality of a
  point with respect of a particular criterion,
\item Braess-like paradoxes: reflects whether an increase of the
  system resource can be detrimental to \emph{all} users concurrently.
\end{itemize}

The rest of this section is organized as follows: after defining these
three fundamental concepts, we study the link between
Pareto-optimality and index optimization, the continuity of
allocations and their monotonicity.

\subsection{Common Definitions}

We recall here the definitions of Pareto optimality, index-optimizing
function and Braess-like paradoxes.

\begin{definition}[Pareto optimality]
  A choice $u\in U$ is said to be Pareto optimal if
  \begin{equation*}
    \forall v\in U,
      \exists i, v_i > u_i \Rightarrow 
      \exists j, v_j < u_j.
  \end{equation*}
  In other words, $u$ is Pareto optimal if it is maximal in $U$ for the
  canonical partial order on $\Rpn$.

  A policy function is said to be \emph{Pareto-optimal} if
  for all $U\in \U, \alpha(U)$ is Pareto-optimal.
\end{definition}

The key idea here is that Pareto optimality is a global notion. Even
in systems that consists of independent elements, the Pareto
optimality cannot be determined on each independent subsystem.  Such
phenomena has been exhibited in~\cite{LT_INFOCOM07}. The considered
system is a master-slave platform in which the master can communicate
with as many slaves as it needs at any time. The master holds a
infinite number of tasks corresponding to $N$ applications, and each
of them can be executed on any slave. The authors study the system at
the Nash equilibrium (each application competing with each other for
both resource and CPU). Although the problems associated with each
machine is independent, the authors show that for any system with one
slave the equilibrium is Pareto optimal, while Pareto inefficiency can
occur in multiple slave systems.

\begin{definition}[$f$-increasing]
  A policy $alpha$ is said to be \emph{$f$-increasing} if $f \circ
  \alpha$ is monotone. Any $f$-optimizing policy is thus
  $f$-increasing.
\end{definition}

\begin{definition}[Common Indexes]~ Many different indexes have been
  proposed in the literature. We present a few ones:\label{comm-index}
  \begin{itemize}
  \item \textbf{Arithmetic mean}: $\sum_i u_i$.
  \item \textbf{Minimum}: $\min_i u_i$.
  \item \textbf{Maximum}: $\max_i u_i$.
  \item \textbf{Geometric Mean}: also called Nash Bargaining Solution or proportional fairness
    $\prod_i u_i$.
  \item \textbf{Harmonic Mean}:
    $\frac{1}{\sum_i 1/u_i}$.
  \item \textbf{Quasi-arithmetic Mean}:
    $f^{-1}(\frac{1}{n}\sum_{i=1}^n f(u_i))$ where $f$ is a strictly
    monotone continuous function on $[0,+\infty]$. The particular case
    where $f$ is defined by $f:x\to x^\delta$ has been widely
    studied~\cite{mo}. The five previous index are particular case of
    this index for particular values of $\delta$ (respectively,
    $1,-\infty, +\infty, 0$ and $-1$).
  \item \textbf{Jain}: $\frac{\left(\sum u_i\right)^2}{\sum u_i^2}$
    (see~\cite{Jain}).
  \item \textbf{Ordered Weighted Averaging}: $OWA(u_1,\dots,u_n)=
    \sum_i w_i.u_{\sigma(i)}$ where $\sigma$ is a permutation such
    that $u_{\sigma(1)}\leq u_{\sigma(2)} \leq\dots\leq
    u_{\sigma(n)}$.
  \end{itemize}
  All these indexes are continuous, however, some of them are not
  strictly monotone.
\end{definition}

\begin{definition}[Braess-paradox]
  A policy function $\alpha$ is said to have Braess-paradoxes it there
  exists $\U_1$ and $\U_2$ such that
  \begin{equation*}
    \U_1 \subset \U_2 \text{ and } \alpha(\U_1) \psucc \alpha(\U_2)
  \end{equation*}
  with $\psucc$ defined as in definition~\ref{canonical}.
  A policy function such that there is no Braess-paradox is called
  \emph{Braess-paradox-free}.
\end{definition}

\subsection{Pareto-optimality and Index Optimization}
\label{sec:pareto-index}
Pareto optimality and monotonicity of the index optimization are
closely related, as illustrated in the following results.

\begin{theorem}
  Let $\alpha$ be an $f$-optimizing policy. If $f$ is strictly
  monotone then $\alpha$ is Pareto-optimal.
\end{theorem}
\begin{proof}
  Suppose that $\alpha$ is not Pareto optimal. Then, there exists $U$
  such that $\alpha(U)$ is not Pareto optimal. Hence, there exists
  $v\in U$ such that $\alpha(U) \prec v$, and hence $f(\alpha(U)) <
  f(v)$, which contradicts the definition of $\alpha(U)$.
\end{proof}

\begin{theorem}
  Let $\alpha$ be an $f$-optimizing policy. If $\alpha$ is
  Pareto-optimal then $f$ is monotone.
\end{theorem}
\begin{proof}
  Suppose that $f$ is not monotone. Then there exists $u \prec v$ such
  that $f(v) \prec f(u)$. Consider $U=\{u,v\}$. As $u \prec v$ and
  $\alpha$ is Pareto-optimal, then $\alpha(U)=v$ which is in
  contradiction with $f(v) \prec f(u)$.
\end{proof}

The Jain index is an example of non-monotone index. The $\min$ and the
$\max$ index are also not strictly monotone, which is why, max-min
fairness or min-max fairness are recursively defined in the
literature.

\subsection{Continuity}

Let us assume that the set of resources is modeled as a compact $R$ of
\Rpn[p] and that utility of users $g$ are continuous functions from
\Rpn[p] to \Rp. Then utility sets $U$ are built with the help of $R$
and $g$.

\begin{equation*}
  U:
  \begin{cases}
    (\H[p],C(\Rpn[p],\Rp)) &\to \H[p] \\
    (R,g) &\mapsto \{g(r)|r\in R\}
  \end{cases}
\end{equation*}

The mapping $U$ being continuous, $\alpha\circ U$ represents the
sensibility of the allocation with respect to resources and utility
functions.  Continuity of the allocation $\alpha$ is thus an essential
feature. Indeed, it ensures that a slight change in the system
resources would not significantly affect the allocation. In
dynamically changing systems, this ensures a certain stability. It
also ensures that a slight error in utility functions does not affect
too much the allocation.

\begin{theorem}
  The Pareto set of a convex utility set is not necessarily compact.

  The function \Pb from \C to \H that associates to $U$ the closure of
  its Pareto set is not continuous.
\end{theorem}
\begin{proof}
  Let us first exhibit a convex utility set whose Pareto set is not
  closed. Let $C=\{(x,y,z)\in \Rpn[3] | x^2+y^2\leq 1, 0\leq z \leq
  1-\frac{1}{2} \frac{x^2-2y^2+1}{1-y}\}$. The set $C$ is depicted
  on Figure~\ref{fig:Pareto_non_closed}. We have
  $\Pb(C)=\{(x,y,1-\frac{1}{2}\frac{x^2-2y^2+1}{1-y}) | x^2+y^2\leq
  1,x+y\geq 1, x>0\}\cup \{(0,1,1)\}$, which is not closed.
  \begin{figure}[ht]
    \centering
    \includegraphics[width=0.8\linewidth]{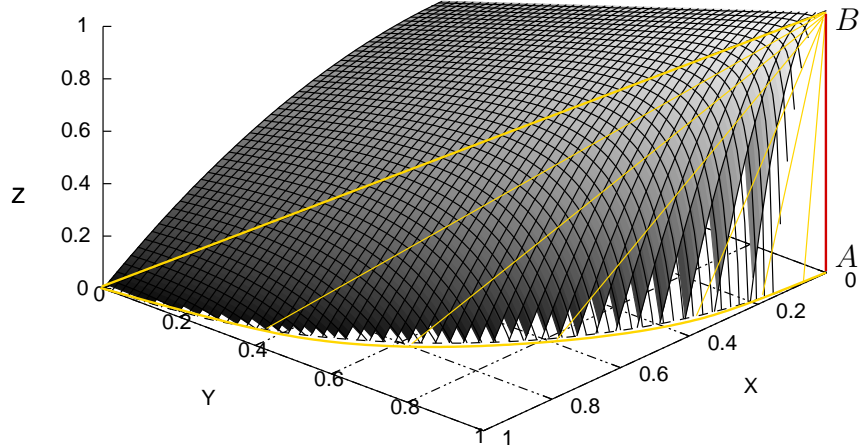}
    \caption{Convex set whose Pareto set is not closed. The segment
      $[A,B[$ does not belong to the Pareto set.}
    \label{fig:Pareto_non_closed}
  \end{figure}

  Let us consider \Pb from \C to \H that associate to $U$ the closure
  of its Pareto set. Figure~\ref{fig:pareto_non_continuous} depicts a
  converging sequence of convex $C_n$ such that $\Pb(C_n)$ does not
  converge to $\Pb(C_{\infty})$.
  \begin{figure}[ht]
    \centering
    \includegraphics[width=.25\linewidth,subfig=1]{pareto_non_continuity.fig}\hfil
    \includegraphics[width=.25\linewidth,subfig=2]{pareto_non_continuity.fig}\hfil
    \includegraphics[width=.25\linewidth,subfig=3]{pareto_non_continuity.fig}
    \caption{\Pb is not continuous.}
    \label{fig:pareto_non_continuous}
  \end{figure}
\end{proof}

\begin{theorem}
  Let $\alpha$ be a general Pareto-optimal policy function. $\alpha$
  is not continuous.
\end{theorem}
\begin{proof}
  We prove that $\alpha$ cannot be continuous with the simple
  instances depicted on Figure~\ref{fig:non_continuous}.  The only
  Pareto-optimal points are $A$ and $B$. Therefore $\alpha$ has to
  choose in the first set between $A$ and $B$. If $A_1$ is chosen,
  then by moving $A_1$ to $A_0$, the choice has to ``jump'' to $B$,
  hence $\alpha$ is not continuous.
  \begin{figure}[ht]
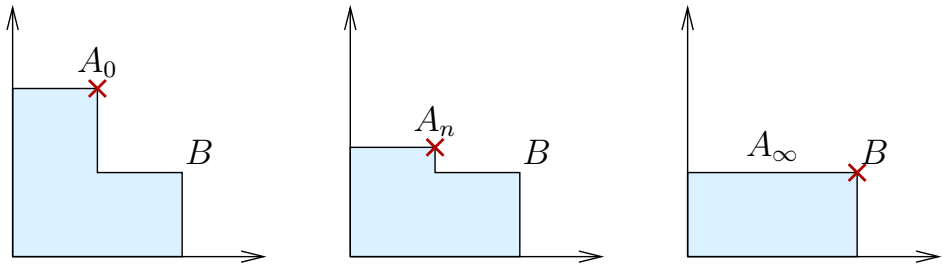

    \centering
    \includegraphics[width=.25\linewidth,subfig=1]{non_continuity.fig}\hfil
    \includegraphics[width=.25\linewidth,subfig=2]{non_continuity.fig}\hfil
    \includegraphics[width=.25\linewidth,subfig=3]{non_continuity.fig}
    \caption{General Pareto-optimal policies are discontinuous: a
      path leading to discontinuity.}
    \label{fig:non_continuous}
  \end{figure}
\end{proof}

\begin{remark}
  There exists continuous and non-continuous convex Pareto-optimal
  policy functions.
\end{remark}
\begin{proof}
  Let us consider a policy function $\alpha$ optimizing the sum of
  utilities. The two convex sets on
  Figure~\ref{fig:convex-non_continuous} show that $\alpha$ is not
  continuous around the set $K=\{(x,y)|x+y\leq 1\}$. This
  discontinuity is due to the fact that many different points of $K$
  simultaneously optimize the sum.
  \begin{figure}[ht]
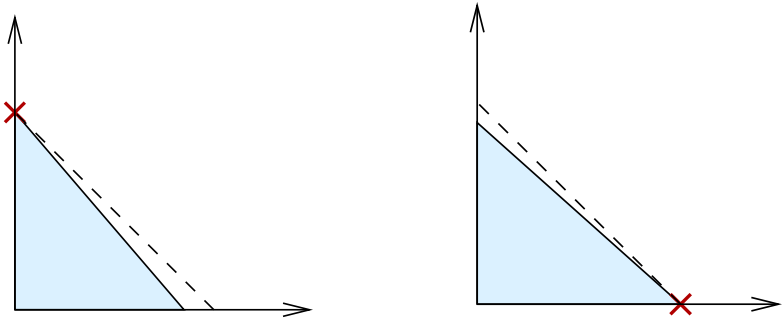

    \centering
    \includegraphics[width=.3\linewidth,subfig=1]{convex_non_continuity.fig}\hfil
    \includegraphics[width=.3\linewidth,subfig=2]{convex_non_continuity.fig}
    \caption{Optimizing $\sum$: a discontinuous convex policy.}
    \label{fig:convex-non_continuous}
  \end{figure}

  The policy function $\alpha$ optimizing the product $\prod$ of
  utilities is continuous though. As $\prod$ is strictly monotone,
  $\alpha$ is Pareto-optimal. Moreover, as for any $c$,
  $I_c=\{x\in\Rpn|\prod x_i\geq c\}$ is strictly convex, for any
  convex, there is a single point optimizing the $\prod$. Let us
  assume by contradiction that $\alpha$ is not continuous at the point
  $C$. Then there exists $C_n$ converging to $C$ and such that
  $x_n=\alpha(C_n)$ converges to $x_{\infty}\neq \alpha(C)$. As our
  sets are compact, there exists a sequence $y_n\in C_n$ such that
  $y_n$ converges to $\alpha(C)$. By definition, we have $\forall n,
  \prod(y_n) \leq \prod(x_n)$. Therefore $\prod(\alpha(C)) \leq
  \prod(x_{\infty})$, which is absurd as $\alpha(C)$ is optimal in $C$
  for $\prod$ and $\alpha(C)\neq x_{\infty}$.
\end{proof}
\subsection{Monotonicity}

We state in this sub-section two results on monotonicity of index and
policy functions. The first one emphasizes that index-functions only
measures a specific characteristic of performance measure, and are
hence not compatible. This explains why allocations that are efficient
(optimizing the arithmetic mean) cannot (in general) also be fair
(optimizing the geometric mean).

The second result states that, even when restricted to convex utility
sets, policy functions cannot be monotone. This infers that even in
Braess-free systems, and increase in the resource can be detrimental
to some users. 

\begin{theorem}\label{monotone1}
  Let $f$ and $g$ be two monotone index functions.  A $g$-optimizing
  policy $\alpha_g$ is $f$-increasing if and only if $\alpha_g$ is
  $f$-optimizing.
\end{theorem}
\begin{proof}
  If $\alpha_g$ is $f$-optimizing, then $\alpha_g$ is clearly
  $f$-increasing.

  Let us assume that $\alpha_g$ is not $f$-optimizing.  We define the
  partial order $\prec_f$ (resp. $\prec_g$) on $\Rpn$ by $x \prec_f y$
  iff $f(x)\leq f(y)$. We have $\prec_f\neq \prec_g$, otherwise
  $\alpha_g$ would be $f$-optimizing. Thus there exists
  $\overline{x_1}$ and $\overline{x_2}$ such that:
  $\overline{x_1}\prec_f \overline{x_2}$ and $\overline{x_2}\prec_g
  \overline{x_1}$. Considering $U=\{x_1\}$ and $U'=\{x_1,x_2\}$, shows
  that $\alpha_g$ is not $f$-increasing.
\end{proof}
In other words, a policy optimizing an index $f$ is always
non-monotone for a \emph{distinct} index $g$.

\begin{theorem}
  Even if convex, policy functions cannot be monotone.
\end{theorem}
\begin{proof}
  Let us consider $\alpha$ a monotone convex policy function and let
  us consider the three following convex sets $U_1=\{(0,1)\}$,
  $U_2=\{(1,0)\}$, and $U_3=\{(x,1-x)|0\leq x\leq 1\}$ (see
  Figure~\ref{fig:convex_non_monotone}).

  \begin{figure}[ht]
    \centering
    \includegraphics[width=.3\linewidth]{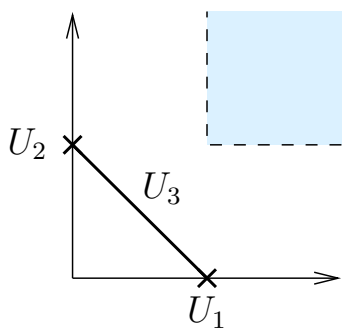}
    \caption{Even convex policy functions cannot be monotone.}
    \label{fig:convex_non_monotone}
  \end{figure}

  We necessarily have $\alpha(U_1)=(0,1)$ and $\alpha(U_2)=(1,0)$. As
  $U_1$ and $U_2$ are subsets of $U_3$, we have $\alpha(U_3) \succeq
  (1,1)$, which is absurd because no such point belongs to $U_3$.
\end{proof}

\subsection{Conclusion}

In this section, we have established the following results:
\begin{itemize}
\item Indexes should be strictly monotone to ensure Pareto-Optimality.
\item Continuity (of allocations) is only possible when considering convex utility sets.
\item It is impossible to ensure that the growth of the utility set
  does not incur the decrease of the utility of some player (\ie
  policy functions cannot be monotone, even when restricting to
  convex utility sets).
\item A policy optimizing a given index $f$ leads to erratic values
  of an other index $g$ when growing utility sets (unless $f$ and $g$
  induce the same optimization).
\end{itemize}
Note that even though being Braess-paradox-free does not lead to bad
properties, it does not give any information on the \emph{efficiency}
of such policies. For example, an allocation $\alpha$ that would be
defined as returning $1/1000$ of the NBS to all users would obviously
be Braess-paradox-free but is very inefficient. This calls for more
quantitative characterization of efficiency.

\section{Quantitative Characterizations}
\label{sec:quant}
How to measure the efficiency of a given policy is still an open
question. Many approaches have been proposed in the literature but we
will see in Section~\ref{sec:qual.discussion} that none of the
previously proposed approach is fully satisfying. We discuss in
particular the most two popular ones: the \emph{Jain
  index}~\cite{Jain} and the \emph{Price of Anarchy}~\cite{papa}. Then
in Section~\ref{sec:qual.topo}, we propose a new metric based on a
more topological point of view and explain how it relates to the
notion of $\epsilon$-approximation~\cite{papa_yannakakis}.


\subsection{Discussion}
\label{sec:qual.discussion}
\subsubsection{Jain index}
The Jain efficiency measure (or Jain index)~\cite{Jain} of a choice $u$
is defined as $\frac{\left(\sum u_i\right)^2}{n\sum u_i^2}$.  The Jain
index is thus the ratio of the first to the second moment of the
choice $u$. Hence, it is a good measure of a choice fairness (as
defined by max-min fairness). The Jain index has many
interesting properties:
\begin{itemize}
\item It is independent of the number of users.
\item It remains unchanged if the utility set is linearly scaled.
\item It is bounded (by $1/n$ and $1$).
\item It is continuous.
\end{itemize}
It can be straight-forwardly adapted to any measure of fairness when
considering the ratio of the first and second moment of $z$ where
for all $i$, $z_i = u_i / v_i$ where $v_i$ is the fair considered
point. Another interpretation of the Jain index is to write it as:
$1/n \sum_i (u_i / u_f)$ where $u_f = (\sum u_i^2) / (\sum u_i)$.
Then each $u_i / u_f$ represents the ratio of the choice with the fair
allocation. The Jain is then the mean of these values. The index is
therefore considered a useful ``distance'' measure to a given fair
point.

The interest of the Jain factor is to determine which users are
discriminated, and which are favored in a given allocation. Transfer of share from
favored to discriminate users always increase the index, while the
opposite reduces it.

\begin{figure*}[htb]
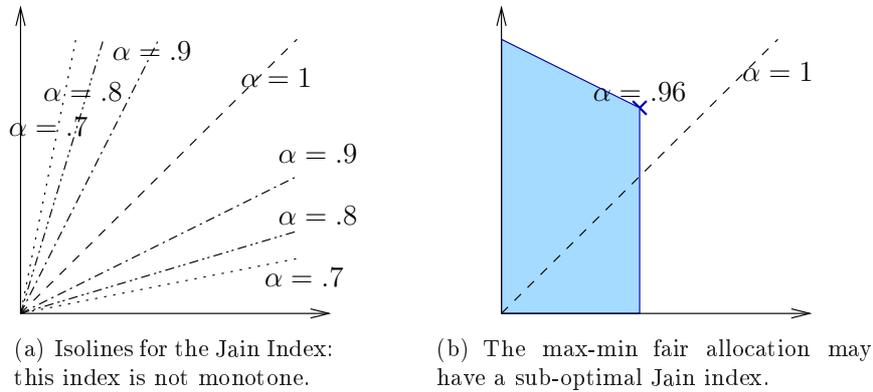

  \centering%
  \subfigure[Isolines for the Jain Index: this index is not
  monotone.]{\label{fig:jain_crappy.a}\includegraphics[width=.3\linewidth]{jain_crappy.fig}}\hfil%
  \subfigure[The max-min fair allocation may have a sub-optimal Jain
  index.]{\label{fig:jain_crappy.b}\qquad\includegraphics[width=.3\linewidth]{jain_crappy2.fig}\qquad}%
  \caption{Highlighting Jain's index flaws.}
  \label{fig:jain_crappy}
\end{figure*}

However, as we have seen in Section~\ref{sec:pareto-index} the Jain
index is non-monotone (see Figure~\ref{fig:jain_crappy.a}), hence
optimal solutions for the Jain index may not be Pareto-optimal. Even
worse, some max-min fair allocations (that are as ``fair'' as possible)
may have sub-optimal Jain index. Such an example is given on
Figure~\ref{fig:jain_crappy.b}.

\subsubsection{Price of Anarchy and Index-Optimizing Based Metrics}
\label{sec:sdf}
\def\SMN{\ensuremath{S_{M,N}}\xspace}

Index-optimizing based metrics are easy to compute, continuous and
generally conserve Pareto-superiority (under some mild conditions). It
is thus natural to select an index $f$ and to try to compare an
allocation to the optimal one for $f$. Papadimitriou~\cite{papa}
introduced the now popular measure ``price of anarchy'' that we will
study in this section.

For a given index $f$, let us consider $\cons[f]$ a $f$-optimizing
policy function. We define the inefficiency $I_f(\beta,U)$ of the
allocation $\beta(U)$ for $f$ as
\begin{align}
  \label{eq:pareto_inefficiency_poa}
  I_f(\beta,U) &= \frac{f(\cons[f](U))}{f(\beta(U))}\geq 1\notag\\
  &= \max_{u\in U} \frac{f(u)}{f(\beta(U))}.
\end{align}
Papadimitriou focuses on the arithmetic mean $\Sigma$ defined by
$\Sigma(u_1,\dots,u_k) = \sum_{k=1}^K u_k$. The price of anarchy
$\phi_\Sigma$ is thus defined as the largest inefficiency:
\begin{equation*}
  \phi_\Sigma(\beta) = \sup_{U\in\U} I_f(\beta,U) = \sup_{U\in\U} 
  \frac{\sum_k \cons[\Sigma](U)_k}{\sum_k \beta(U)_k}
\end{equation*}
In other words, $\phi_{\Sigma}(\beta)$ is the \emph{approximation
  ratio} of $\beta$ for the objective function $\Sigma$. This measure
is very popular and rather easy to understand. However, we will see
that it may not reflect what people have in mind when speaking about
``price of anarchy''.

\begin{figure}
  \centering
  \includegraphics[width=.6\linewidth]{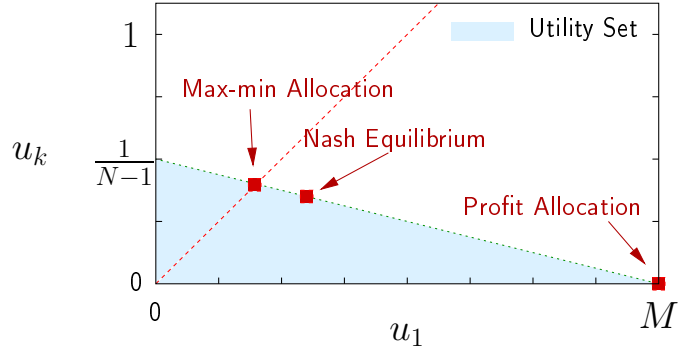}
  \caption{Utility set and allocations for $\SMN$ ($N=3$,$M=2$), with
    $u_2=\dots=u_N$.}
  \label{fig:Pareto}
\end{figure}

Consider the utility set $\SMN=\{u\in\Rpn[N] | u_1/M+\sum_{k=1}^{N}
u_k\leq 1\}$ depicted in Fig~\ref{fig:Pareto}. As the roles of the
$u_k$, $k\geq 2$ are symmetric, we can freely assume that
$u_2=\dots=u_N$ for metrical index-optimizing policies.
\begin{remark}
  This example was taken from the master-slave scheduling problem
  of~\cite{LT_INFOCOM07}.
\end{remark}
It is then easy to compute the following index optimizing allocation:
\begin{itemize}
\item $\cons[\Sigma](\SMN)=(M, 0,\dots,0)$ corresponds to the allocation
  optimizing the average utility;
\item $\cons[\min](\SMN)=\left(\frac{1}{N-1+1/M},\dots,
    \frac{1}{N-1+1/M}\right)$ corresponds to the max-min fair
  allocation~\cite{equite};
\item $\cons[\Pi](\SMN)=\left(\frac{M}{N},\frac{1}{N},\dots,
    \frac{1}{N}\right)$ corresponds to the proportionally fair
  allocation which is a particular Nash Bargaining
  Solution~\cite{equite}.
\end{itemize}
Note that, \cons[\Sigma], \cons[\min], and \cons[\Pi] are Pareto
optimal by definition. One can easily compute the price of anarchy of
the Nash Bargaining solution:
\begin{equation*}
  I_\Sigma(\cons[\Pi],\SMN) = \frac{M}{\frac{M}{N}+\frac{N-1}{N}}
  \xrightarrow[M\rightarrow\infty]{} N.
\end{equation*}
The price of anarchy is therefore unbounded. However, the fact that
this allocation is Pareto-optimal and has interesting properties of
fairness (it corresponds to a Nash Bargaining Solution~\cite{equite})
questions the relevance of the \emph{price of anarchy} notion as a
Pareto efficiency measure.

Likewise, the inefficiency of the max-min fair allocation is
equivalent to $M$ for large values of $M$ (as opposed to $K$ for the
non-cooperative equilibrium). It can hence be unbounded even for
bounded number of applications and machines. This seems even more
surprising as such points generally result from complex cooperations
and are hence Pareto optimal. These remarks raise once more the
question of the measure of Pareto inefficiency.

These are due to the fact that a policy optimizing an index $f$ is
always non-monotone for a distinct index $g$ (from
Theorem~\ref{monotone1}). Hence any policy (including Pareto optimal
ones) optimizing a distinct index from the arithmetic mean will
experience a bad price of anarchy. Note that the previous problems are
not specific to the efficiency measure arithmetic mean. The same kind
of behavior can be exhibited when using the $\min$ or the product of
the throughputs for instance.

That is why we think that Pareto inefficiency should be measured as the
\emph{distance} to the Pareto border and not to a specific point.

\subsubsection{Selfishness Degradation Factor}

To quantify the degradation of Braess-like Paradoxes (the degree of
Paradox), Kameda~\cite{kameda06} introduced the Pareto-compa\-rison of
$\alpha$ and $\beta$ as $\rho(\alpha,\beta) = \min_k
\frac{\alpha_k}{\beta_k}$.  Therefore, $\alpha$ is strictly superior
to $\beta$ iff $\rho(\alpha,\beta)>1$. Intuitively $\rho$ represents
the performance degradation between $\alpha$ and $\beta$. Using this
definition, the following
definition of Pareto inefficiency, named Selfishness Degradation
Factor (SDF), was proposed \cite{LT_INFOCOM07}:
\begin{equation}
  \label{eq:pareto_inefficiency_sdf}
  \ISDF(\beta,U) = \max_{u\in U} \rho(u,\beta(U)) =
  \max_{u\in U} \min_k \frac{u_k}{\beta(U)_k}
\end{equation}
Therefore $\beta(U)$ is Pareto inefficient as soon as
$\ISDF(\beta,U)>1$ and the larger $\ISDF(\beta,U)$, the more
inefficient the allocation. 
\def\bplus{{\boxplus}}
\begin{figure*}
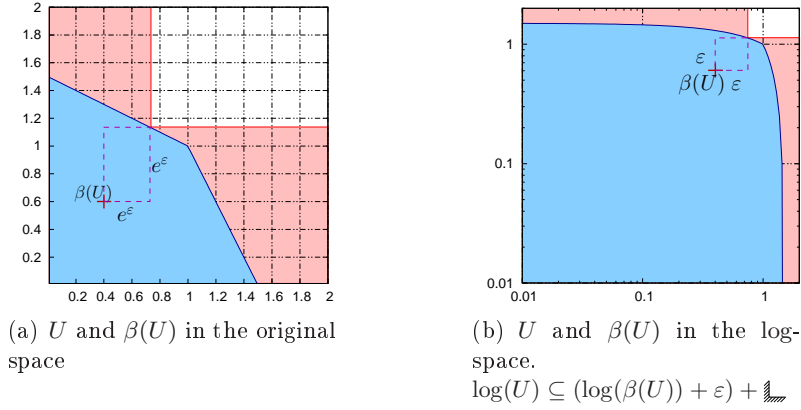

  \centering%
  \subfigure[$U$ and $\beta(U)$ in the original
  space]{\label{fig:sdf.normal}%
    \includegraphics[width=.3\linewidth,subfig=3]{Sdf1.fig}}\hfil%
  \subfigure[$U$ and $\beta(U)$ in the $\log$-space.\newline
  \usebox{\rsbox}]{\label{fig:sdf.log}%
    \includegraphics[width=.3\linewidth,subfig=4]{Sdf1.fig}}
  \caption{Inefficiency for the selfishness degradation factor:
    $\log(U) \subset (\log(\beta(U))+\epsilon)$}
  \label{fig:sdf}
\end{figure*}
\begin{lemma}

  Let us define $\lowerarea=\{x\in \Rn| \exists k: x_k\leq 0\}$. We
  denote by $a\bplus\lowerarea=\{x\in \Rn| \exists k: x_k\leq
  a_k\}$.

  \begin{equation*}
    \log(\ISDF(\beta,U))\leq \epsilon \Leftrightarrow \log(U)\subseteq
    (\log(\beta(U))+\epsilon) \bplus \lowerarea
  \end{equation*}
\end{lemma}
\begin{proof}
  \begin{align*}
    \log(\ISDF(\beta,U))\leq \epsilon & \Leftrightarrow
    \max_{u\in U} \min_k \log(\frac{u_k}{\beta(U)_k}) \leq \epsilon\\
    & \Leftrightarrow \forall u\in U, \exists k, \log(u_k) -
    log(\beta(U)_k) \leq \epsilon \\
    & \Leftrightarrow \forall u\in U, \exists k, \log(u_k) \leq
    log(\beta(U)_k) + \epsilon \\
    & \Leftrightarrow \forall u\in U, \log(u) \in
    (\log(\beta(U))+\epsilon) \bplus \lowerarea \\
    & \Leftrightarrow \log(U) \subseteq (\log(\beta(U))+\epsilon) \bplus
    \lowerarea
  \end{align*}
\end{proof}
Figure~\ref{fig:sdf} depicts a graphical interpretation of this
inefficiency measure.  As illustrated by the previous lemma, this
inefficiency seems to measure how much $\beta(U)$ should be increased
so that it is not dominated by any other points in $U$. Therefore,
$\log(\ISDF(\beta,U))$ somehow measures the distance in the log-space
from $\beta(U)$ to the Pareto set. However, as we will see in the next
section, this definition holds only because of the very specific shape
of the set $\U$ used in this example.

Anyway, the \emph{selfishness degradation factor} can, as in
section~\ref{sec:sdf}, be defined from this inefficiency measure:
\begin{equation*}
  \phi = \sup_{U\in\U} \ISDF(\beta,U) = \sup_{U\in\U} \max_{u \in U}
  \min_k \frac{u_k}{\beta(U)_k}.
\end{equation*}
A system (\eg queuing network, transportation network, load-balancing,
...) that would be such that the Nash equilibria are always Pareto
optimal would have a selfishness degradation factor equal to one. The
selfishness degradation factor may however be unbounded on systems
where non-cooperative equilibria are particularly inefficient. The
relevance of this definition is corroborated by the fact that
$\epsilon$-approximations of Pareto-sets defined by Yannakakis and
Papadimitriou~\cite{papa_yannakakis} have a degradation factor of
$\exp(\epsilon)\simeq 1+\epsilon$.

\subsection{A Topological Point of View}
\label{sec:qual.topo}
In this section, we go back to the inefficiency measure introduced in
the previous section and show that such a measure can be properly
defined only when referring to the whole Pareto set. Indeed, what we
are interested in is in fact some kind of distance of a point to the
Pareto set. As researchers are used to look at factors when evaluating
the performance of an algorithm, this distance to the Pareto set
should be measured in the log space. As we have seen in the previous
section, the inefficiency measure for the selfishness degradation
factor is closely related to the distance to the Pareto set. More
precisely, we prove that being close to the Pareto set implies a small
measure of inefficiency. However, the converse is true only when the
utility set has some particular properties.

The distance from $\beta(U)$ to the closure of the Pareto set $\Pb(U)$
in the log-space is equal to:
\begin{equation*}
  d_{\infty}(\log(\beta(U),\log(\Pb(U))) = \min_{u\in\Pb(u)} \max_k
  |\log(\beta(U)_k) - \log(u_k)|
\end{equation*}
Therefore, we can define
\begin{align}
  \I(\beta,U)&=\exp(d_{\infty}(\log(\beta(U),\log(\Pb(U))) \notag\\
  &= \min_{u\in\Pb(u)} \max_k \max\left(\frac{\beta(U)_k}{u_k},
    \frac{u_k}{\beta(U)_k}\right)
\end{align}

Let us recall the classical expansion definition:
\begin{equation*}
X\oplus
a=\{y|d(x,y)\leq a, \text{ for some } x\in X\}
\end{equation*}
This definition can be easily expanded as:
\begin{align*}
  X\otimes a &= \exp(\log(X)\oplus\log(a)) \\ &=  
  \{y|\exp(d(\log(x),\log(y))\leq a \text{ for some } x\in X\} 
\end{align*}

\begin{definition}[$\epsilon$-approximation]
  \cite{papa_yannakakis} defines an $\epsilon$-approxima\-tion of
  $\Pb(U)$ as a set of points $S$ such that for all $u\in U$ there
  exists some $s\in S$ such that $\forall k: u_k\leq
  (1+\epsilon)s_k$.
\end{definition}
With the previous notations, it is easy to see that:

\begin{theorem}
  $S\subseteq U$ is an $\epsilon$-approximation of $\Pb(U)$ iff
  $\Pb(U)\subseteq S\otimes \exp(\epsilon)$.
\end{theorem}

\begin{figure*}[htb]
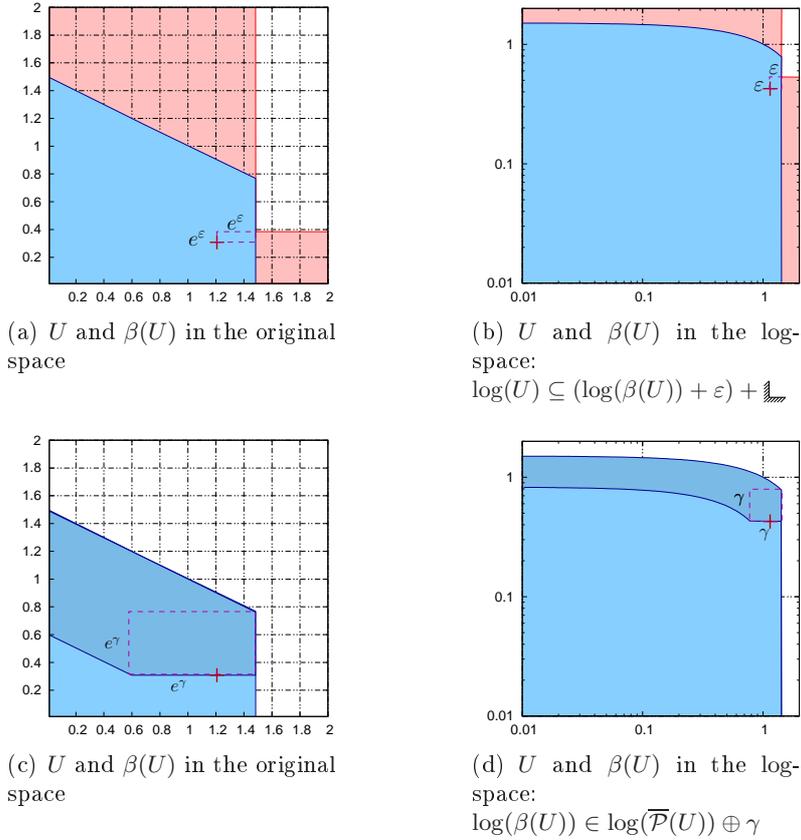

  \label{fig:topo}
  \centering%
  \subfigure[$U$ and $\beta(U)$ in the original
  space]{\label{fig:topo.normal}%
    \includegraphics[width=.3\linewidth,subfig=3]{Sdf2.fig}}\hfil%
  \subfigure[$U$ and $\beta(U)$ in the $\log$-space: \newline
  \usebox{\rsbox}]{\label{fig:topo.log}%
    \includegraphics[width=.3\linewidth,subfig=4]{Sdf2.fig}}\hfil%
  \\
  \subfigure[$U$ and $\beta(U)$ in the original 
  space]{\label{fig:topo.normal.expension}%
    \includegraphics[width=.3\linewidth,subfig=5]{Sdf2.fig}}\hfil%
  \subfigure[$U$ and $\beta(U)$ in the $\log$-space:\newline
  $\log(\beta(U))\in \log(\Pb(U))\oplus
  \gamma$]{\label{fig:topo.log.expension}%
    \includegraphics[width=.3\linewidth,subfig=6]{Sdf2.fig}}\hfil%
  \caption{Distance to the Pareto set}
\end{figure*}

Figure~\ref{fig:topo.log.expension} depicts the expansion of
$\log(\Pb(U))$ by $\epsilon$ so that it contains $\log(\beta(U))$. It
is easy to show that:
\begin{lemma}
  $\I(\beta,U)\leq \exp(\epsilon) \Leftrightarrow \beta(U)\in
  \Pb(U)\otimes \exp(\epsilon)$.
\end{lemma}
In other words, $\I(\beta,U)\leq \exp(\epsilon)$ iff $\beta(U)$ is no
farther than $\epsilon$ from $\Pb(U)$ in the log space.

When comparing the definitions of $I_\Sigma$, \ISDF and \I, the latest
may seem harder to compute as it relies on $\P(U)$. However, what we
are interested in is measuring the distance to the Pareto set and no
index-based inefficiency measure can reflect this distance. Then can
only reflect a particular property of the allocation such as
fairness. Note that in mono-criteria situations, it is natural to
compare a solution to an intractable optimal solution, generally using
approximations or lower bounds. Therefore, similar approaches should
be used in multi-criteria settings to compute \I. This inefficiency
measure is thus a natural extension of the classical mono-criteria
performance ratio.

The previous definition should thus be used in the general case, even
though in a some particular situations, the SDF definition is
sufficient.

\section{Conclusion}
\label{sec:conclusion}

In this paper, we have addressed the question of how to properly
measure efficiency of allocations, may they be obtained as the result
of some index-function optimization or some general policy. We have
shown a number of results, both at qualitative and
quantitative level. In particular, we have shown that:
\begin{itemize}
\item Monotonicity is the link between index-optimization and Pareto
  optimality.
\item When utilities are continuous with the system's resources,
  solution allocations can be continuous in the resources only when
  the utility sets are convex.
\item Even with Braess-free allocations, there always exists instances
  where resource increase is detrimental to at least one user.
\item A policy optimizing a given index leads to erratic values for
  another index when utility sets grow.
\item Both the Jain index and the price of anarchy have flaws as
  measures of the inefficiency of an equilibria.
\item A correct general inefficiency measure can be defined and is
  based on the log space as the distance of a point to the Pareto
  border.
\end{itemize}
We believe that these results can serve as a general theoretical
milestones to any researcher aiming at analyzing the performance of an
allocation in a specific problem.

\bibliographystyle{abbrv}
\bibliography{biblio}

\end{document}